\documentstyle[fleqn]{article}

\begin{document}
\date{}
\author{V.N.Gorbachev, A.I.Trubilko}
\title{Quantum teleportation of EPR pair 
by three-particle entanglement 
}

\maketitle
\begin{abstract}
Teleportation of an EPR pair using triplet in state of 
the Horne-Greenberger-Zeilinger form to two receivers is considered.
It needs a three-particle basis for joint measurement.
By contrast the one qubit teleportation the required basis 
is not maximally entangled. It consists of the states 
corresponding to the maximally entanglement of two particles 
only. Using outcomes of measurement both receivers can recover
an unknown EPR state however one of them can not do it 
separately. Teleportation of the N-particle entanglement is 
discussed.
\end{abstract}
\section{Introduction}
The teleportation process, proposed by Bennett et al 
\cite{1}, allows to transmit an unknown state of 
a quantum system from a sender, traditionally named Alice, 
to receiver, or Bob, both are spatially separated.
For teleporting of a two-state particle, or qubit, 
it needs an EPR pair and a usual communication channel.
The large number of versions using two-particle 
entanglement has been considered \cite{2}.
Quantum teleportation of the photon polarized \cite{4} and   
a single coherent mode of field \cite{40} has been demonstrated in optical 
experiments.

As a source of EPR pair light can be used, particularly
the light of the optical parametric oscillator or down 
conversion as in \cite{40}. 
However the physical nature of the particles may be 
different, for instance one can choose the EPR-correlated atoms 
realized experimentally in \cite{41}. Indeed, 
the particles of different nature are introduced 
in the scheme called the inter-space teleportation \cite{42},
where quantum state is transferred, for example, between the atom 
and light \cite{43}.   

In this work we consider teleportation of an entangled pair  
to two distant parties Bob and Claire with the use of the 
triplet in the Greenberger-Horne-Zeilinger state (GHZ). Indeed
the GHZ triplet has been realized experimentally \cite{5}.
The main problem is to find the three-particle projection 
basis for 
a joint measurement. By contrast the single qubit state 
teleportation, the maximally entangled basis does not 
accomplish the task. The obtained basis consist of a set of
the three-particle projection operators with the 
maximally entanglement of two particles only.
Measuring allows both receivers to recover 
an unknown state of EPR pair, but each of them 
can not do it  separately. As it has been shown in ref. \cite{6},
where teleportation of a single qubit using 
the GHZ triplet has been considered,  
only one of the receivers and not both can recovered an unknown state.
Our results are generalized for the 
teleportation of the N-particle entanglement as the EPR-nplet.

Our work is organized as follows. In section 2 the main 
features of the teleportation of a single qubit are given.
The initial states of the entangled pair and triplet are discussed 
in section 3. In section 4 
the basis for the joint measurement is found.
The teleportation protocol and network are presented in section 5, 
where the results are generalized for the N- particle entanglement.

\section{Teleportation}
The teleportation of an unknown quantum state between 
two parties spatially separated, Alice and Bob, includes the following 
steeps \cite{1}.
Let Alice has a two level system or qubit prepared in an unknown 
state 
\begin{equation}
|\psi_{1}\rangle=\alpha|0\rangle +\beta|1\rangle
\label{01}
\end{equation}
where $|\alpha|^{2}+|\beta|^{2}=1$. 
Let Alice and Bob share a maximally entangled EPR pair
$|\Psi_{23}\rangle =(|01\rangle +|10\rangle )/\sqrt{2}$, so that qubit 2 
is for Alice and qubit 3 is for Bob.
First, Alice performs a joint measurement of qubits 1 and 2 in the 
Bell basis consisting of four projectors 
$\Pi_{k}=|\pi_{k}\rangle \langle \pi_{k}|$, $k=1,\dots 4$, 
$|\pi_{1}\rangle =|\Phi ^{+}_{12}\rangle$,
$|\pi_{2}\rangle =|\Phi ^{-}_{12}\rangle$,
$|\pi_{3}\rangle =|\Psi ^{+}_{12}\rangle$,
$|\pi_{4}\rangle =|\Psi ^{-}_{12}\rangle$,
where the Bell states are the maximally entanglement 
of two particles  
\begin{equation}
|\Phi ^{\pm}\rangle =\frac{1}{\sqrt{2}}(|00\rangle \pm |11\rangle )
\label{1}
\end{equation}
\begin{equation}
|\Psi ^{\pm}\rangle =\frac{1}{\sqrt{2}}(|01\rangle \pm |10\rangle )
\label{2}
\end{equation}

As result of the joint measurement, the density operator of the 
combined system 
$\rho=|\psi_{1}\rangle \langle \psi_{1}|
\otimes |\Psi_{23}\rangle \langle \Psi_{23}|$, 
to be defined in the three-particle Hilbert space
$H_{1}\otimes H_{2}\otimes H_{3}$, 
is projected into one of four Bell states.
Two point are important in this procedure, first, the k-th outcome 
depends not on $\psi_{1}$
second, the reduced density matrix of the qubit 3 
$\rho_{3}(k)=Sp_{12}\{\Pi_{k}\rho\Pi_{k}^{\dagger}\}$ 
and unknown state both are connected by the unitary 
transformation $U_{k}$
\begin{equation}
\rho_{3}(k)=U_{k}\tilde \rho_{1} U_{k}^{\dagger}
\label{3}
\end{equation}
where
$\tilde \rho_{1}$ is the density operator of $H_{3}$, 
that is the counterpart state
$\rho_{1}=|\psi_{1}\rangle \langle \psi_{1}|$, $U_{k}$ is the 
set of the Pauli matrices 
$U_{1}=\sigma_{x}, U_{2}=-i\sigma_{y}, U_{3}=1, 
U_{4}=\sigma_{z}$.  
Finally Alice sends the outcomes of her measurement to Bob 
who performs on his qubit 3 one of four unitary operations, 
corresponding Alice' message and has his qubit in the 
original state    
$\psi_{1}$. Teleportation is achieved.

\section{Initial states}

To teleport an EPR pair it needs a maximally entanglement of 
three particles. From this fact let consider what initial 
states would be used.

The wave function of an entangled pair can be chosen  as
\begin{equation}
|\Psi_{12}\rangle = \alpha |00\rangle +\beta |11\rangle
\label{0001}
\end{equation}
where $|\alpha|^{2}+|\beta|^{2}=1$, or in the form of EPR-pair  
\begin{equation}
|\Psi_{EPR}\rangle = \alpha |01\rangle +\beta |10\rangle
\label{0002}
\end{equation}
It is possible to point eight states where three particles are maximally entangled. 
They are
\begin{eqnarray}
(|000\rangle \pm |111\rangle )/\sqrt{2},&\quad& (|001\rangle \pm |110\rangle )/\sqrt{2},
\nonumber
\\
(|010\rangle \pm |101\rangle )/\sqrt{2},&\quad& (|100\rangle \pm |011\rangle )/\sqrt{2}
\label{103}
\end{eqnarray}
From the presented set of the initial states of particles without 
loss of generality we choose
(\ref{0002}) and triplet in the form of $GHZ$
\begin{equation}
|\Psi_{GHZ}\rangle =\frac{1}{\sqrt{2}}(|000\rangle +|111\rangle )
\label{102}
\end{equation} 
Now we shall consider the combined system prepared initially 
in the state 
\begin{equation}
|\Psi\rangle =|\Psi_{EPR}\rangle \otimes |\Psi_{GHZ}\rangle
\label{100}
\end{equation}

\begin{figure}
\unitlength=1.00mm
\special{em:linewidth 0.4pt}
\linethickness{0.4pt}
\begin{picture}(92.67,44.67)
\put(20.67,41.33){\line(1,0){16.00}}
\put(20.67,35.00){\line(1,0){15.67}}
\put(20.33,28.33){\line(1,0){16.33}}
\put(37.33,41.33){\circle{2.00}}
\put(37.00,35.00){\circle{2.00}}
\put(37.33,28.33){\circle{2.00}}
\put(32.67,24.33){\framebox(9.67,20.33)[cc]{}}
\put(20.33,21.00){\line(1,0){36.00}}
\put(56.33,21.00){\line(0,0){0.00}}
\put(56.67,17.33){\framebox(7.33,6.33)[cc]{B}}
\put(20.00,14.33){\line(1,0){51.67}}
\put(71.33,10.67){\framebox(7.33,6.33)[cc]{C}}
\put(64.33,21.00){\line(1,0){28.00}}
\put(78.67,14.33){\line(1,0){14.00}}
\put(42.67,34.33){\rule{18.33\unitlength}{1.00\unitlength}}
\put(61.33,33.67){\line(0,-1){10.00}}
\put(61.33,33.67){\line(5,-2){14.00}}
\put(75.33,28.00){\line(0,-1){11.00}}
\put(11.67,21.00){\line(1,0){11.33}}
\put(11.33,21.33){\line(4,3){9.33}}
\put(11.33,21.00){\line(4,-3){9.00}}
\put(11.67,38.00){\line(5,2){8.67}}
\put(12.00,38.00){\line(3,-1){9.33}}
\put(61.33,34.67){\circle*{2.67}}
\put(11.00,38.00){\circle{2.00}}
\put(10.33,21.33){\circle{2.00}}
\put(10.00,44.00){\makebox(0,0)[cc]{$\Psi_{EPR}$}}
\put(9.67,28.67){\makebox(0,0)[cc]{GHZ}}
\put(27.00,44.33){\makebox(0,0)[cc]{1}}
\put(27.00,38.33){\makebox(0,0)[cc]{2}}
\put(27.00,31.33){\makebox(0,0)[cc]{3}}
\put(27.00,24.67){\makebox(0,0)[cc]{4}}
\put(26.67,17.33){\makebox(0,0)[cc]{5}}
\put(46.00,44.67){\makebox(0,0)[cc]{A}}
\put(87.67,27.00){\makebox(0,0)[cc]{$\Psi_{EPR}$}}
\end{picture}
\caption{Teleportation of EPR pair of qubits 1 and 2 using GHZ triplet. Alice, Bob and Claire
share qubits 3,4 and 5 of GHZ. Alice sends outcome of a joint measurement to 
Bob and Claire who recover an unknown EPR-state.}
\end{figure}
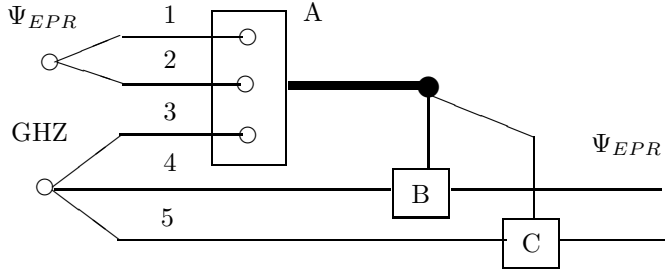
The scheme to teleport an unknown state of EPR pair of 
qubit 1 and 2 with the use of the GHZ triplet of 
qubit 3,4 and 5 is presented in fig. 1. 
Here three parties spatially separated, Alice and two receivers 
Bob and Claire share the GHZ particles 3,4 and 5. 
Alice sends outcomes of her joint measurement of qubits 1,2 
and 3 to both receivers by classical channel. To perform the joint 
measurement it needs 
eight projection operators to form a complete set
on which the initial  wave function $|\Psi\rangle $
can be decomposed. The choice of such basis is the main moment in the solution 
of the problem. 

\section{The projection basis}

It would be possible to imagine that the set of the projection 
operators $\Pi_{k}$ for joint measurement will consist of the maximally
 entangled states (\ref{103}). Let denote this basis as
$\pi_{(123)}$. However one can find that it is not true. The reason is 
that in series expansion of the initial wave function $|\Psi\rangle $ 
its projections into four vectors 
$(|010\rangle \pm |101\rangle)/\sqrt{2}$,  
$(|100\rangle \pm |011\rangle)/\sqrt{2}$ of the basis $\pi_{(123)}$
are equal to zero.
It is impossible to recover unknown state of EPR pair by the such 
outcomes using unitary transformation. Therefore the maximally entangled tree- particle basis does not
solve the task. 

The basis required turns out to be composed from the states 
where only two particle are maximally entangled, say 1,3 or 1,2.
However the pair entangled is only the necessary condition.

To consider realization of operators $\Pi_{k}$ we introduce classification
where one of tag will be number of the particles to be maximally entangled, 
say two or three in our case. As all complete set of vectors are 
connected among themselves by unitary transformation 
one can take an initial basis. Let the initial basis be $\pi_{123}$
\begin{equation}
|\pi_{123}\rangle =|ijk\rangle \quad i,j,k =0,1
\label{p123}
\end{equation}
where each of eight elements is the state of three 
independent or non-correlated particles. Any element of the other 
basis can be presented by a linear superposition of
$s\leq 8$ vectors of the set $\pi_{123}$. Further let suppose the 
number s be common for the given basis and we use it for 
classification. So it can be introduced the set 
$\pi_{1(23)}(s)$ consisting of the maximally entanglement of 
two particles 2 and 3. 
For $s=2$ it has the form 
\begin{equation}
|\pi_{1(23)}(2)\rangle =\{|i\rangle |\Phi^{\pm}_{23}\rangle ; 
|i\rangle |\Psi^{\pm}_{23}\rangle \} \quad i=0,1
\label{p}
\end{equation}
where each of eight vectors, for example
$|0\rangle |\Phi^{\pm}_{23}\rangle =(|000\rangle \pm |011\rangle )/\sqrt{2}$, 
is presented by two elements of 
$\pi_{123}$. 
For the case $s=4$
\begin{equation}
|\pi_{1(23)}(4)\rangle =\{
|\pi_{1}^{\pm}\rangle |\Phi^{\pm}_{23}\rangle ;
|\pi_{1}^{\pm}\rangle |\Psi^{\pm}_{23}\rangle \}
\label{pp}
\end{equation}
where pair of vectors generating a complete single-particle set
looks like
\begin{equation}
|\pi_{1}^{\pm}\rangle =\frac{1}{\sqrt{2}}
(|0\rangle \pm \exp(i\varphi)|1\rangle )
\label{ppp}
\end{equation}

The sets presented here are complete and orthogonal, however the basis
$\pi_{1(23)}(2)$ does not solve the problem. The reason is that 
the outcomes of the joint measurement depend on the wave function to be 
teleported so that the unitary transformation 
that Bob and Claire have to perform 
at their qubits will depend on an unknown state. For base 
$\pi_{(12)3}(4)$ the situation is similar to 
$\pi_{(123)}(2)$, where half of projections of $\Psi$ into the 
basis states is equal to zero.

For telepoting  EPR pair two sets are useful for which $s=4$. 
There are $\pi_{1(23)}(4)$ ore  $\pi_{(13)2}(4)$, 
where the particles 2,3 
or 1,3 are maximally entangled. The structure of the initial 
state, projection basis $\pi_{1(23)}(4)$ and the total wave function 
are presented in fig 2.  
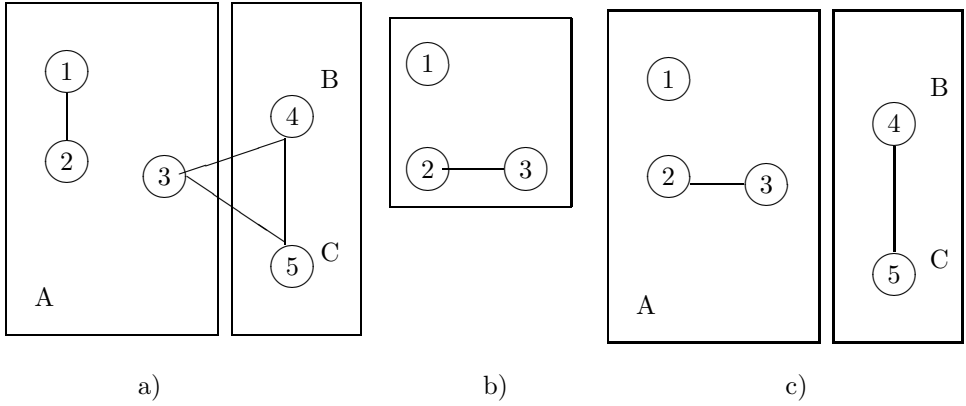
\begin{figure}[ht]
\unitlength=1.00mm
\special{em:linewidth 0.4pt}
\linethickness{0.4pt}
\begin{picture}(130.00,52.00)
\put(11.00,43.00){\circle{5.66}}
\put(11.00,31.00){\circle{5.66}}
\put(24.00,29.00){\circle{5.66}}
\put(41.00,37.00){\circle{5.66}}
\put(41.00,17.00){\circle{5.66}}
\put(11.00,43.00){\makebox(0,0)[cc]{1}}
\put(11.00,31.00){\makebox(0,0)[cc]{2}}
\put(24.00,29.00){\makebox(0,0)[cc]{3}}
\put(41.00,37.00){\makebox(0,0)[cc]{4}}
\put(41.00,17.00){\makebox(0,0)[cc]{5}}
\put(3.00,8.00){\framebox(28.00,44.00)[cc]{}}
\put(33.00,8.00){\framebox(17.00,44.00)[cc]{}}
\put(40.00,20.00){\line(0,1){14.00}}
\put(40.00,34.00){\line(-3,-1){14.00}}
\put(59.00,44.00){\circle{5.66}}
\put(59.00,30.00){\circle{5.66}}
\put(72.00,30.00){\circle{5.66}}
\put(59.00,44.00){\makebox(0,0)[cc]{1}}
\put(59.00,30.00){\makebox(0,0)[cc]{2}}
\put(72.00,30.00){\makebox(0,0)[cc]{3}}
\put(91.00,42.00){\circle{5.66}}
\put(91.00,29.00){\circle{5.66}}
\put(104.00,28.00){\circle{5.66}}
\put(121.00,36.00){\circle{5.66}}
\put(121.00,16.00){\circle{5.66}}
\put(91.00,42.00){\makebox(0,0)[cc]{1}}
\put(91.00,29.00){\makebox(0,0)[cc]{2}}
\put(104.00,28.00){\makebox(0,0)[cc]{3}}
\put(121.00,36.00){\makebox(0,0)[cc]{4}}
\put(121.00,16.00){\makebox(0,0)[cc]{5}}
\put(83.00,7.00){\framebox(28.00,44.00)[cc]{}}
\put(113.00,7.00){\framebox(17.00,44.00)[cc]{}}
\put(121.00,19.00){\line(0,1){14.00}}
\put(8.00,13.00){\makebox(0,0)[cc]{A}}
\put(46.00,42.00){\makebox(0,0)[cc]{B}}
\put(46.00,19.00){\makebox(0,0)[cc]{C}}
\put(88.00,12.00){\makebox(0,0)[cc]{A}}
\put(127.00,41.00){\makebox(0,0)[cc]{B}}
\put(127.00,18.00){\makebox(0,0)[cc]{C}}
\put(27.00,29.00){\line(3,-2){13.00}}
\put(22.00,1.00){\makebox(0,0)[cc]{a)}}
\put(68.00,1.00){\makebox(0,0)[cc]{b)}}
\put(108.00,1.00){\makebox(0,0)[cc]{c)}}
\put(54.00,25.00){\framebox(24.00,25.00)[cc]{}}
\put(11.00,40.00){\line(0,-1){6.00}}
\put(61.00,30.00){\line(1,0){8.00}}
\put(94.00,28.00){\line(1,0){7.00}}
\end{picture}
\caption{The entanglement structure of the states. 
a) The initial state, where
qubits of EPR pair 1,2 and qubits 3,4,5 of GHZ are entangled.
b) Projection basis $\pi_{1(23)}(4)$. 
c) The state after measuring.}
\end{figure}
 
\section{Teleportation of EPR pair}
Using the obtained set $\pi_{1(23)}$, where we put 
$\varphi =0$, the initial wave function can be rewritten as
\begin{eqnarray}
|\Psi\rangle&=&
|\pi_{1}^{+}\rangle |\Phi^{+}_{23}\rangle |1\rangle +
|\pi_{1}^{+}\rangle |\Phi^{-}_{23}\rangle |2\rangle 
\nonumber
\\ 
&+&
|\pi_{1}^{-}\rangle |\Phi^{+}_{23}\rangle |3\rangle +
|\pi_{1}^{-}\rangle |\Phi^{-}_{23}\rangle |4\rangle 
\nonumber
\\
&+&
|\pi_{1}^{+}\rangle |\Psi^{+}_{23}\rangle |5\rangle +
|\pi_{1}^{+}\rangle |\Psi^{-}_{23}\rangle |6\rangle 
\nonumber
\\
&+&
|\pi_{1}^{-}\rangle |\Psi^{+}_{23}\rangle |7\rangle +
|\pi_{1}^{-}\rangle |\Psi^{-}_{23}\rangle |8\rangle 
\label{201}
\end{eqnarray}
where
\begin{eqnarray}
|1,2\rangle &=& \beta|00\rangle \pm \alpha |11\rangle 
\nonumber
\\
|3,4\rangle &=&-(\beta|00\rangle \mp \alpha |11\rangle  )
\nonumber
\\
|5,6\rangle &=&\beta|11\rangle \pm \alpha |00\rangle 
\nonumber
\\
|7,8\rangle &=&-(\beta|11\rangle \mp \alpha |00\rangle  )
\label{203}
\end{eqnarray}
that for each outcome the reduced density matrix of qubit 4 and 5 
It follows from equation (\ref{203}) 
$\rho_{45}(k)=Sp_{123}\{\Pi_{k}|\Psi\rangle \langle \Psi|\Pi_{k}\}$ 
is connected to the density matrix of EPR pair by unitary transformation
\begin{equation}
\rho_{45}(k)=U_{k}|\tilde\Psi_{EPR}\rangle \langle \tilde\Psi_{EPR}|
U^{\dagger}_{k}
\quad k=1, \dots 8
\label{204}
\end{equation}
where 
$\tilde\Psi_{EPR}$ is the wave function of Hilbert space
$H_{4}\otimes H_{5}$ and counterpart of $\Psi_{EPR}$.
The unitary operator from (\ref{204}) has the form
$U_{1}=\sigma_{x4}\otimes I_{5}$,
$U_{2}=-U_{3}=i\sigma_{y4}\otimes I_{5}$, 
$U_{4}=-U_{1}$,
$U_{5}=I_{4}\otimes \sigma_{x5}$, 
$U_{6}=-U_{7}= I_{4}\otimes (-i\sigma_{y5})$,
$U_{8}=-U_{5}$. 
where the Pauli operators $\sigma_{\gamma j}$
$\gamma =x,y,z$
and identity $I_{j}$  
affect the particle $j=4,5$.

Teleportation of an unknown EPR state can be reached by the following 
protocol:
\begin{enumerate}
\item  
Alice performs the joint measurement of qubits 1,2 and 3 
in basis $\pi_{1(23)}(4)$ and sends her outcomes to Bob and Claire. 
\item 
For outcomes $k=1 - 4$ 
Bob have to rotate his qubit by local operations
$\sigma_{x}$, $i\sigma_{y}$, $-i\sigma_{y}$, $-\sigma_{x}$
and Claire does nothing. As result Bob and Claire has EPR pair
in the state $\Psi_{EPR}$.
\item 
To recover an unknown EPR state for outcomes $k=5 - 6$
Bob does nothing  and Claire performs unitary transformation 
$\sigma_{x}$, $-i\sigma_{y}$, $i\sigma_{y}$,
$-\sigma_{x}$ on her qubit.
\end{enumerate}
In the presented protocol in half of cases only one of receivers 
affects on his particle while another acts on his particle by 
unity operator or does nothing. This version is not unique 
because unknown state can be recovered by different way. 
For instance, the wave vector 
$|2\rangle $ from (\ref{203}) 
can be obtained by two ways 
\begin{equation}
\beta |00\rangle -\alpha|11\rangle 
=i\sigma_{y4}\otimes I_{5}
|\tilde\Psi_{EPR}\rangle 
=\sigma_{x4}\otimes \sigma_{z5}
|\tilde\Psi_{EPR}\rangle 
\label{205}
\end{equation}
The expression (\ref{205}) means that for outcome  
$k=2$ both receivers Bob and Claire should simultaneously affect 
on their qubits (as in the above protocol) by unitary operations  
$\sigma_{x4}$ and $\sigma_{z5}$ (instead of 
$i\sigma_{y4}$ and identity operator).
These differences however do not change the result. The 
main feature of the teleportation procedure considered here is 
presence of two receivers which one can not accomplish the 
task separately.
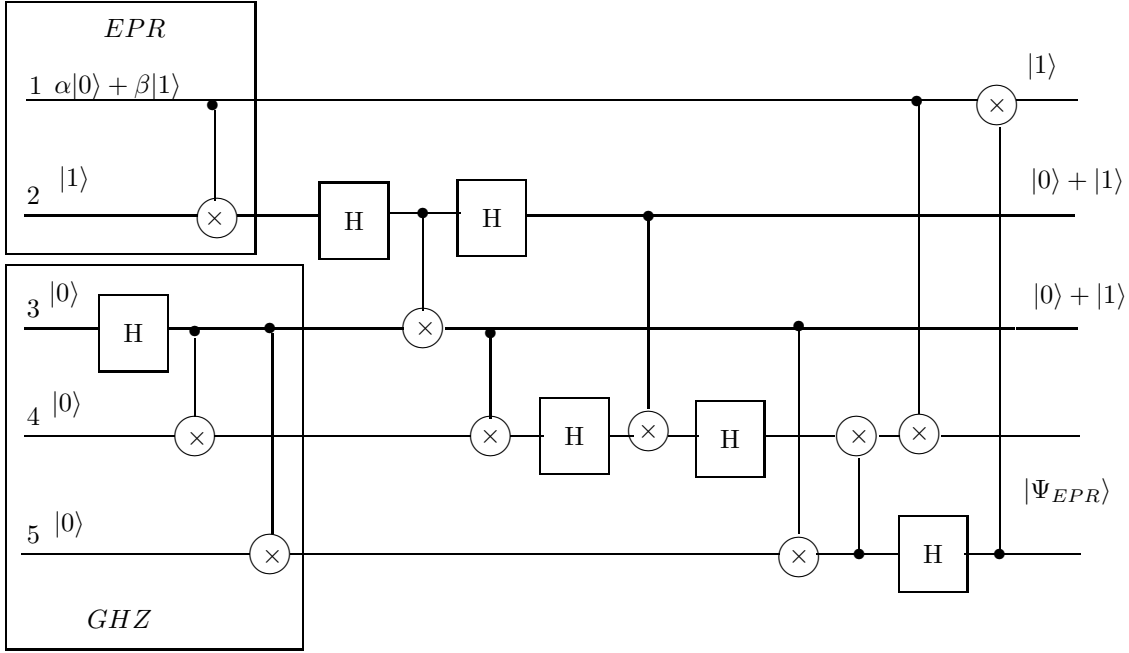
\begin{figure}
\unitlength=1.00mm
\special{em:linewidth 0.4pt}
\linethickness{0.4pt}
\begin{picture}(147.67,91.00)
\put(7.67,78.00){\line(1,0){20.33}}
\put(136.67,77.33){\circle{5.37}}
\put(32.33,77.34){\circle*{1.33}}
\put(32.99,62.34){\circle{5.37}}
\put(7.33,62.67){\line(1,0){18.33}}
\put(46.67,57.00){\framebox(9.00,10.00)[cc]{H}}
\put(55.67,63.00){\line(1,0){9.67}}
\put(65.00,57.33){\framebox(9.00,10.00)[cc]{H}}
\put(7.33,47.67){\line(1,0){9.67}}
\put(17.33,42.00){\framebox(9.00,10.00)[cc]{H}}
\put(60.33,63.00){\circle*{1.33}}
\put(60.33,47.67){\circle{5.37}}
\put(26.33,47.67){\line(1,0){31.33}}
\put(63.33,47.67){\line(1,0){75.67}}
\put(74.00,62.67){\line(1,0){65.33}}
\put(32.67,76.67){\line(0,-1){12.00}}
\put(60.33,62.67){\line(0,-1){12.67}}
\put(30.00,47.33){\circle*{1.33}}
\put(40.00,47.67){\circle*{1.33}}
\put(30.00,33.33){\circle{5.37}}
\put(7.33,33.33){\line(1,0){20.00}}
\put(40.00,17.67){\circle{5.37}}
\put(7.00,17.67){\line(1,0){30.33}}
\put(30.00,46.33){\line(0,-1){10.33}}
\put(40.33,47.00){\line(0,-1){26.67}}
\put(69.33,47.00){\circle*{1.33}}
\put(69.33,33.33){\circle{5.37}}
\put(32.67,33.33){\line(1,0){34.00}}
\put(76.00,28.33){\framebox(9.00,10.00)[cc]{H}}
\put(90.33,34.00){\circle{5.37}}
\put(123.67,12.67){\framebox(9.00,10.00)[cc]{H}}
\put(118.00,33.33){\circle{5.37}}
\put(126.33,33.67){\circle{5.37}}
\put(137.00,17.67){\circle*{1.33}}
\put(118.33,17.67){\circle*{1.33}}
\put(96.67,28.00){\framebox(9.00,10.00)[cc]{H}}
\put(110.33,48.00){\circle*{1.33}}
\put(110.33,17.33){\circle{5.37}}
\put(128.00,78.00){\line(1,0){6.00}}
\put(139.33,78.00){\line(1,0){8.00}}
\put(139.33,62.67){\line(1,0){7.67}}
\put(139.33,47.67){\line(1,0){8.00}}
\put(147.33,47.67){\line(0,0){0.00}}
\put(72.00,33.33){\line(1,0){4.33}}
\put(85.00,33.33){\line(1,0){3.33}}
\put(93.00,33.33){\line(1,0){4.00}}
\put(105.67,33.33){\line(1,0){9.33}}
\put(121.00,33.33){\line(1,0){2.67}}
\put(129.33,33.33){\line(1,0){18.33}}
\put(42.67,17.67){\line(1,0){65.00}}
\put(112.67,17.67){\line(1,0){5.33}}
\put(132.33,17.67){\line(1,0){15.33}}
\put(119.00,17.67){\line(1,0){4.67}}
\put(126.00,78.00){\circle*{1.33}}
\put(90.33,62.67){\circle*{1.33}}
\put(69.33,47.00){\line(0,-1){11.33}}
\put(90.33,62.33){\line(0,-1){25.33}}
\put(110.33,47.67){\line(0,-1){27.33}}
\put(118.33,30.33){\line(0,-1){13.00}}
\put(126.33,77.67){\line(0,-1){41.33}}
\put(137.00,74.33){\line(0,-1){56.67}}
\put(30.33,33.00){\makebox(0,0)[cc]{$\times$}}
\put(40.33,17.33){\makebox(0,0)[cc]{$\times$}}
\put(60.67,47.67){\makebox(0,0)[cc]{$\times$}}
\put(69.33,33.33){\makebox(0,0)[cc]{$\times$}}
\put(90.33,34.00){\makebox(0,0)[cc]{$\times$}}
\put(110.33,17.33){\makebox(0,0)[cc]{$\times$}}
\put(118.33,33.33){\makebox(0,0)[cc]{$\times$}}
\put(126.33,33.67){\makebox(0,0)[cc]{$\times$}}
\put(136.67,77.33){\makebox(0,0)[cc]{$\times$}}
\put(8.00,80.00){\makebox(0,0)[lc]{$1$}}
\put(11.33,79.00){\makebox(0,0)[lb]{$\alpha|0\rangle+\beta|1\rangle$}}
\put(27.67,78.00){\line(1,0){98.67}}
\put(43.00,62.67){\line(1,0){3.33}}
\put(7.67,65.33){\makebox(0,0)[lc]{2}}
\put(7.67,50.33){\makebox(0,0)[lc]{3}}
\put(7.67,36.00){\makebox(0,0)[lc]{4}}
\put(7.67,20.33){\makebox(0,0)[lc]{5}}
\put(14.00,66.67){\makebox(0,0)[cb]{$|1\rangle$}}
\put(12.67,51.33){\makebox(0,0)[cb]{$|0\rangle$}}
\put(13.00,36.67){\makebox(0,0)[cb]{$|0\rangle$}}
\put(13.33,20.67){\makebox(0,0)[cb]{$|0\rangle$}}
\put(142.67,81.67){\makebox(0,0)[cb]{$|1\rangle$}}
\put(147.34,66.34){\makebox(0,0)[cb]{$|0\rangle+|1\rangle$}}
\put(147.67,51.01){\makebox(0,0)[cb]{$|0\rangle+|1\rangle$}}
\put(24.67,62.67){\line(1,0){5.67}}
\put(35.67,62.67){\line(1,0){11.00}}
\put(32.67,62.33){\makebox(0,0)[cc]{$\times$}}
\put(5.00,57.67){\framebox(33.00,33.33)[cc]{}}
\put(22.00,87.67){\makebox(0,0)[cc]{$EPR$}}
\put(5.00,5.00){\framebox(39.33,51.00)[cc]{}}
\put(20.00,9.00){\makebox(0,0)[cc]{$GHZ$}}
\put(126.33,78.00){\line(1,0){3.00}}
\put(146.00,26.00){\makebox(0,0)[cc]{$|\Psi_{EPR}\rangle$}}
\end{picture}
\caption{Network for teleportation of EPR pair}
\end{figure}

The network presented in fig. 3 illustrates the teleportation 
procedure of an EPR state. It is built similarly to the  
one-particle teleportation \cite{7}, 
and includes set of logical operations C-NOT (controlled- not) and 
Hadamard transformation H. In the unit $EPR$ operation C-NOT $C_{12}$
acting qubit 1,2 produces the entanglement of particles 1,2 of the 
form of EPR state.
$C_{12}$ flips the second qubit (target) if and only if 
the first (control) is logical 1.  
The unit  $GHZ$ prepares three-particle entanglement by the Hadamard 
transformation H acting qubit 3 
($H|0\rangle=(|0\rangle +|1\rangle )/\sqrt{2}$, 
$H|1\rangle=(|0\rangle -|1\rangle )/\sqrt{2})$ 
and two operations c-NOT 
$C_{34}, C_{35}$. On the end of the scheme the joint state of qubits 
4 and 5 being independent of others is turned out to be 
$\Psi_{EPR}$. Indeed, the above network can be used for teleporting 
the entangled pair of the form (\ref{0001}).

Consider the more general case of teleportation of the N-particle 
entanglement as EPR-nplet
\begin{equation}
|\Psi_{N}\rangle =\alpha |0\rangle ^{N}
+\beta |1\rangle ^{N}
\label{210}
\end{equation}
using $N+1$ qubits in the maximally entangled state as GHZ
\begin{equation}
|\Psi_{(N+1)}\rangle = \frac{1}{\sqrt{2}}
\left(|0\rangle ^{N+1} +
|1\rangle ^{N+1}\right)
\label{211}
\end{equation} 
where $|i\rangle ^{N}= |i\rangle \otimes \dots|i\rangle$, 
$i=0,1$.
In this procedure, that includes 
2N+1 qubits, a sender Alice and N receivers share the entanglement 
of the form (\ref{211}).
The combined wave function defined in the Hilbert space 
$H_{1}\otimes \dots H_{2N+1}$ 
is product
$|\Psi\rangle =|\Psi_{N}\rangle \otimes |\Psi_{(N+1)}\rangle $. 

For joint measuring of
N+1 ($1,2,\dots N,N+1$) particles it needs a complete set of 
$2^{N+1}$ projectors describing states of any maximally 
entangled pair $M,N+1$, $M=1,\dots N$. 
For $M=N$ 
the required basis has the form
\begin{equation}
\pi_{1,...N-1(N,N+1)}(s)=
\{|\pi_{1,...N-1}\rangle \otimes |\Phi^{\pm}_{N,N+1}\rangle ;
  |\pi_{1,...N-1}\rangle \otimes |\Psi^{\pm}_{N,N+1}\rangle \}
\label{212}
\end{equation}
where the particles
$N,N+1$ are entangled,  
$|\pi_{1,...N-1}\rangle $ are the vectors from 
$H_{1}\otimes \dots H_{N-1}$. As we note before the 
presence of the entangled pair is only 
the necessary condition for the required basis. The sufficient 
condition is magnitude of parameter s which one together with vectors 
$\pi_{1,...N-1}$ 
can be obtained by expanding the combined wave function over 
set of (\ref{212}). It can be written as
\begin{eqnarray}
|\Psi \rangle &=&
\{ P_{N-1} \alpha |0\rangle ^{N}\pm Q_{N-1}\beta |1\rangle ^{N}\}
|\pi_{1,...N-1}\rangle |\Phi^{\pm}_{N,N+1}\rangle
\nonumber
\\
&+&
\{ P_{N-1}\alpha |1\rangle ^{N}\pm Q_{N-1}\beta |0\rangle ^{N}\}
|\pi_{1,...N-1}\rangle |\Psi^{\pm}_{N,N+1}\rangle
\label{213}
\end{eqnarray}
where
\begin{equation}
P_{N-1}=\langle \pi_{1,...N-1}|0\rangle ^{N-1}
\qquad
Q_{N-1}=\langle \pi_{1,...N-1}|1\rangle ^{N-1}
\label{214}
\end{equation}
Process teleportation needs the following condition  
\begin{equation}
P_{N-1}\neq 0, Q_{N-1}\neq 0
\label{215}
\end{equation}
It means that all terms of the series expansion of $\Psi$ 
have to involve a linear superposition of
$\alpha |i\rangle ^{N}$ and $\beta|j\rangle ^{N}$, $i\neq j=0,1$, 
which one can be retrieved from 
(\ref{210}) by unitary transformation not dependent from an unknown 
state. The following set of vectors obeys 
(\ref{215})
\begin{equation}
|\pi_{1,...N-1}\rangle =\{|\pi^{\pm}_{1}\rangle ^{N-1}\}
\label{216}
\end{equation}
where $\pi^{\pm}_{1}$ is the one-particle set defined by
(\ref{ppp}). It can be easily established, noting that the set 
(\ref{216}) consists of 
$2^{N-1}$ elements each of which contains two terms 
$|i\rangle ^{N-1}, i=0,1$. 

For the obtained basis defined by 
(\ref{212}) and (\ref{216})
the value $s$ is equal to $2^{N}$. Note, that all cases with 
$s<2^{N}$, where there are bases inclusive more than one pair of 
the entangled particles (two pairs or triplet) does not accomplish 
the task.  

As result, {\em for teleporting an $N$-particle entangled state as EPR -nplet 
using the $N+1$ particle entanglement it needs the set of the
projection operators with one pair of the maximally entangled particles. 
Each element of this set has to be presented by $2^{N}$ 
vectors corresponding the $N+1$-independent particle state}.


\begin{thebibliography}{99}
\bibitem{1}
C.H.Bennet, G.Brasard, C.Crepeau, R.Jozsa, A.Peres, W.K.Wootters.
Phys.Rev.Lett. {\bf 70}, 1895 (1993).
\bibitem{2}
L.Vaidman. Phys.Rev. A {\bf49}, 1473, (1994).\\
S.L.Braunstein, H.J.Kimble. Phys.Rev.Lett. {\bf80}, 869 (1998)\\
S.N. Molotkov. Rus. JETF Lett.,{\bf68}, 248 (1998).\\
P.van Loock, S.L. Braunstein. Broadband teleportation,
e-print quant/ph 9902030.\\
R.E.S.Polkinghome, T.C.Ralph. Entanglement swapping using 
continuous variables. e-print quant/ph 9906066.
\bibitem{4}
D.Bouwmeester, J-W Pan, M.Mattle, H.Weinfurter, A.Zielinger.
Nature, 390, 575, (1997).\\
D.Boschi, S.Branca, F.De Martini, L.Hardy, S.Popescu.
Phys. Rev. Lett. {\bf80}, 1121, (1998).
\bibitem{40}
A.Furusawa, J.L.Sorensen, S.L.Braunstein,
C.A.Fuchs, H.J.Kimble, E.S.Polzik. Science, 282, 706, (1998).
\bibitem{41}
A.Kusmich, K.Notmer, E.S.Polzik. Phys.Rev.Lett., {\bf79}, 4782, (1997).//
J.L.Sorensen, J.Hold, E.S.Polzik. Phys.Rev.Lett. {\bf80}, 3487, (1998).
\bibitem{42}
C.S.Maierle, D.A.Lidar, R.A.Harris. Teleporting Superposition of 
Chiral Amplitudes, e-print quant/ph 9807020. 
\bibitem{43}
A.S.Parkins, H.J.Kimble. Quantum state transfer between motion and 
light, e-print quant/ph 9904054. 
\bibitem{5}
D.Bouwmeester, J.Pan, M.Daniell, H.Weinfurter, A.Zeilinger.
Observation of three photon Greenberger - Horne - Zeilinger 
entanglement, e-print quant/ph 9810035.\\
R.J.Nelson,D.G.Cory, S.Lloyd. Experimental demonstration of 
Greenberger - Horne - Zeilinger correlatons using nuclear 
magnetic resonance, e-print quant/ph 9905028. 
\bibitem{6}
A.Karlson, M.Bourennane. Phys.Rev. A {\bf58}, 4394, (1998).
\bibitem{7}
G.Brassard. Teleportation as a quantum computation, 
e-print quant/ph 9605035.\\
C.A.Adami, N.J.Cerf. Quantum Computation with Linear Optics, 
e-print quant/ph 9806048.\\
S.L.Brounstein. Phys. Rev. A {\bf53}, 1900, (1996).

\end{thebibliography}
\end{document}